\documentclass[conference]{IEEEtran}
\IEEEoverridecommandlockouts
\usepackage{cite}
\usepackage{amsmath,amssymb,amsfonts}
\usepackage{algorithmic}
\usepackage{graphicx}
\usepackage{textcomp}
\usepackage{xcolor}
\usepackage{multirow}
\usepackage{float}
\usepackage{subfig}

\def\BibTeX{{\rm B\kern-.05em{\sc i\kern-.025em b}\kern-.08em
    T\kern-.1667em\lower.7ex\hbox{E}\kern-.125emX}}
\begin{document}

\title{First CE Matters: On the Importance of Long Term Properties on Memory Failure Prediction}

\author{\IEEEauthorblockN{Jasmin Bogatinovski, Odej Kao}
\IEEEauthorblockA{\textit{Technical University Berlin}, 
Berlin, Germany \\
\{jasmin.bogatinovski, odej.kao\}@tu-berlin.de}
\and
\IEEEauthorblockN{Qiao Yu, Jorge Cardoso} 
\IEEEauthorblockA{\textit{Huawei Munich Research},
Munich, Germany \\
University of Coimbra, CISUC, DEI, Coimbra, Portugal \\
jorge.cardoso@huawei.com}
}

\IEEEoverridecommandlockouts
\IEEEpubid{\makebox[\columnwidth]{IEEE BigData 978-1-6654-8045-1/22/\$31.00 \copyright2022 IEEE \hfill} \hspace{\columnsep}\makebox[\columnwidth]{ }}

\maketitle

\begin{abstract}
Dynamic random access memory failures are a threat to the reliability of data centres as they lead to data loss and system crashes. Timely predictions of memory failures allow for taking preventive measures such as server migration and memory replacement. Thereby, memory failure prediction prevents failures from externalizing, and it is a vital task to improve system reliability. In this paper, we revisited the problem of memory failure prediction. We analyzed the correctable errors (CEs) from hardware logs as indicators for a degraded memory state. As memories do not always work with full occupancy, access to faulty memory parts is time distributed. Following this intuition, we observed that important properties for memory failure prediction are distributed through long time intervals. In contrast, related studies, to fit practical constraints, frequently only analyze the CEs from the last fixed-size time interval while ignoring the predating information. Motivated by the observed discrepancy, we study the impact of including the overall (long-range) CE evolution and propose novel features that are calculated incrementally to preserve long-range properties. By coupling the extracted features with machine learning methods, we learn a predictive model to anticipate upcoming failures three hours in advance while improving the average relative precision and recall for 21\% and 19\% accordingly. We evaluated our methodology on real-world memory failures from the server fleet of a large cloud provider, justifying its validity and practicality.  
\end{abstract}


\section{Introduction}
Dynamic random access memory errors are omnipresent failure types in data centres. A memory error is an event which results in the wrong reading of the logical state of one or multiple bits from how they were last written. An error in a single bit corrupts the stored information and affects the ongoing computation, sometimes leading to system crashes. The root causes of DIMM errors are diverse, ranging from hardware corruption of the memory arrays, up to bit-flips due to electromagnetic influence (e.g., cosmic ray strikes)~\cite{errortypes}. As a preventive strategy for DIMM reliability, the manufacturers integrate different error correcting codes (ECC) onto the chip for single (e.g., parity checking)~\cite{SEC} or multi-bit error correction (e.g., ChipKill~\cite{ChipKill}). While useful in a limited set of circumstances (e.g., SEC corrects a single bit), once the errors exceed the assumptions on the ECC algorithms (e.g., the error occurs in multiple bits), the memory failure externalises and frequently results in a system crash. Therefore, relying on error correction codes as a preventive strategy is insufficient for reliable DIMM behaviour. To reduce the effect of the DIMM failures alternative strategies are needed. 

A possible strategy to amortize the effect of DIMM failures is to predict when the memory will fail. By a sufficiently large prediction interval, the correct prediction enables the triggering of preventive measures (e.g., server migration). Therefore, the failure will not be externalised. This makes memory failure prediction an important task for system reliability. The key requirement for memory prediction is the availability of data that encapsulates the properties of a degraded DIMM. Two common data sources are used, i.e., 1) hardware error logs and 2) system-level metrics~\cite{facebookmemoryfailureprediction}. Hardware error logs record the correctable errors generated from the DIMM's ECC chips. These errors are referred to as \textit{correctable errors}. Different studies show that the accumulated repetition of CE on nearby memory locations is correlated with memory failures~\cite{stefanovici2012cosmic,facebookmemoryfailureprediction}. Therefore, DIMM characterization through CEs is useful to model DIMM degradation. In contrast, memory-related system-level metrics are time series points showing the memory utilization (e.g., byte read/write). Although system-level metrics are reported as useful~\cite{memoryfailuresystemmetrics,bogatinovskiMipro2019}, the CEs are generally recognized to better capture the memory degradation and are more frequently used for failure prediction~\cite{facebookmemoryfailureprediction}. 

\begin{figure*}[!ht]
\centering
\subfloat{\includegraphics[width=\columnwidth]{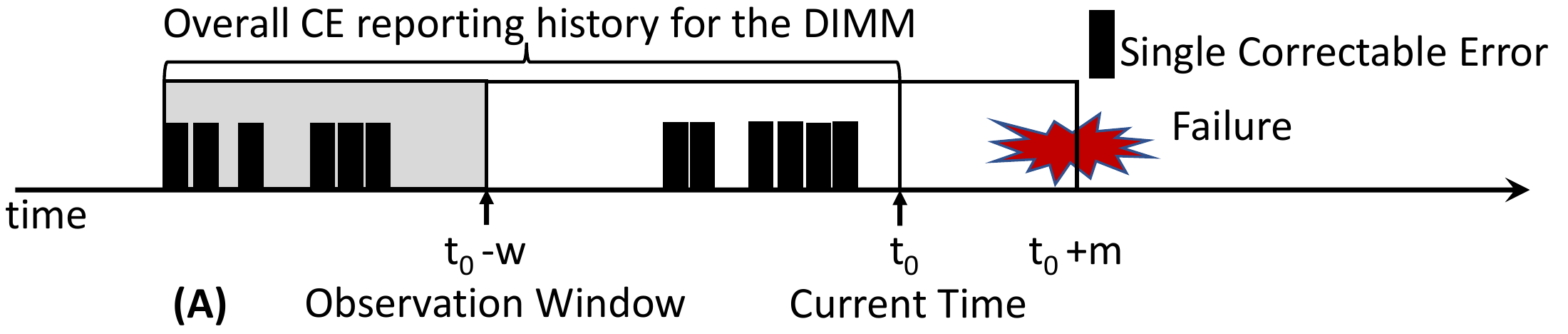}%
\label{timeintervals}}
\hfil
\subfloat{\includegraphics[width=\columnwidth]{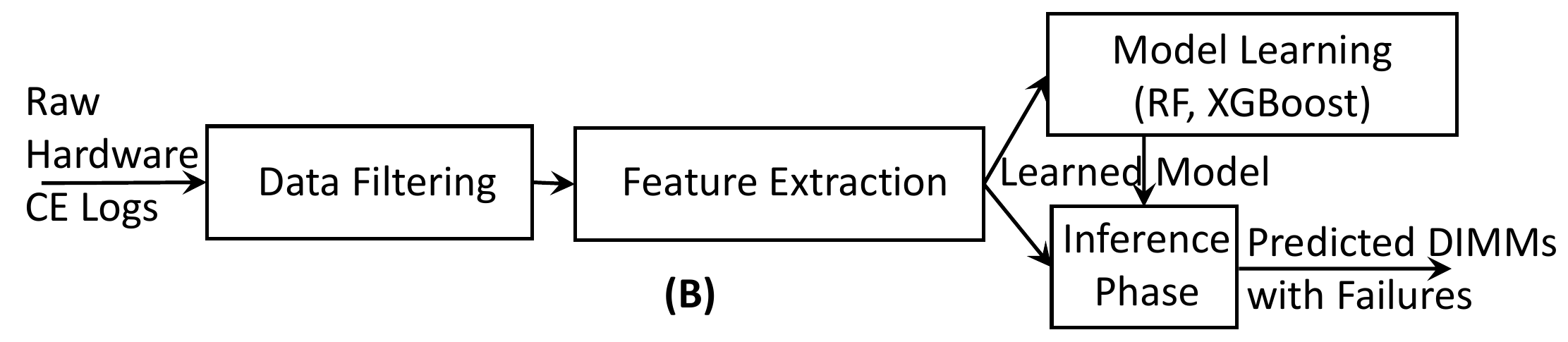}%
\label{pipeline}}
\caption{(A) The Memory Failure Prediction Problem; (B) Model learning pipeline;}
\end{figure*}

In this paper, we revisit the problem of memory failure prediction by using CEs. Owning to practical requirements (e.g., unavailable infrastructure for long-term storage of the overall CE log history), existing approaches introduce an observation window with a fixed time duration as part of the algorithm design. Studies typically consider a fixed window of, e.g., one or two weeks~\cite{memoryfailuresystemmetrics, IBM2018, paper3}. However, as the utilization of the servers has a stochastic component, it is often the case that the servers will be accessed time apart at distant intervals. Furthermore, as memories on a single server do not always work with full occupancy, the access of the faulty memory parts (and CE generation thereof) is time-distributed as well. Therefore, by using fixed observation windows, one may expect that important details from CE history may be missed. Based on this intuition, we examine the impact of considering the overall CEs DIMM history instead of using an observation window of fixed size. We used our observations to propose a set of novel features that can be incrementally calculated while preserving the long-term temporal CEs dependencies of a degraded DIMM not needing to store all the data. By pairing the calculated features with machine learning methods, we learn a predictive model that can anticipate memory failures three hours in advance, outperforming operational practices. Our evaluation is performed on \textbf{real-world memory failure data} collected from the server fleet of a large cloud provider, justifying the validity and practicality of our approach. 

\section{Background}~\label{background}
\vspace{-6mm}
\subsection{System Memory Organization}
The memory system of a server has a hierarchical organization. A single DIMM (dual in-line memory module) is a computer working memory type. The DIMM stores each data bit in a separate memory cell and is used by the CPU for  different operations (e.g., \textit{add}, \textit{write}). The DIMM is connected with the \textit{integrated memory controller} (IMC) within the CPU via a \textit{memory channel}. To increase the DIMM performance, a DIMM is usually composed of multiple integrated \textit{chips} that are organized in \textit{ranks} (set of connected chips). Each chip is further composed from arrays of transistors identifiable by \textit{rows} and \textit{columns}. The intersection between a row and column is called \textit{cell}. Each cell stores one bit. The DIMM has also additional circuitry that enables effective memory operations, and potentially ECC mechanisms for CE event generation. The CE events can be read out from dedicated CPU register set groups (e.g., the Machine Check Exception (MCE) register set for Intel x86) generating the CE logs. Opposed to the properties of the software logs, like semantics~\cite{bogatinovskiSCC2022}, or sequences~\cite{bogatinovskiCCGRID2022}, the CE logs contain information about the place in the memory where the CE occurred (e.g., specific cell, bank, row, etc). 

\subsection{Problem Formulation}
Let $D_i$ is a DIMM that reported a CE log, denoted by $l(t_j)$ at a certain point in time $t_j$, and let $w$ and $m$ denote an observation window length and a prediction time length interval, accordingly. The goal of memory failure prediction is to map $F: \phi(l_{(t_0-w:t_0)}) \mapsto B_m$, where $t_0$ is the current observation moment and $\phi(l_{(t_0-w:t_0)})$ is a representation function $\phi: L_i \mapsto \mathcal{R}^d$ that maps the set of observed correctable errors $L_i$ for DIMM $D_i$ from the time period ($t_0-w$, $t_0$) into a certain representation space $\mathcal{R}^d$, where $d$ is representation size. The set $B_m \in \{0, 1\}$ denotes if the DIMM $i$ will fail in $m$ time units $(1)$ or not $(0)$. 

\begin{figure*}[!t]
\centering
\includegraphics[width=0.7\textwidth]{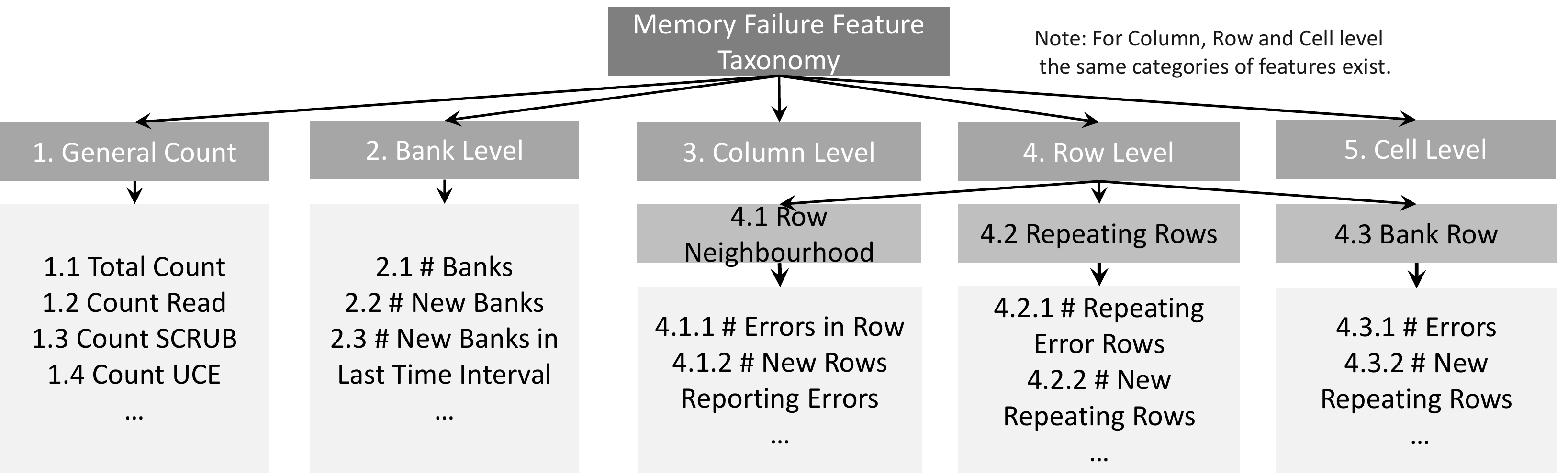}
\caption{Feature Taxonomy; Based on the hierarchical DIMM organization.}
\label{featuretaxonomy}
\end{figure*}

\textbf{Motivation}. In previous works~\cite{memoryfailuresystemmetrics, IBM2018, paper3}, the observation window $w$ used for feature extraction is usually fixed, e.g., two weeks. Figure~\ref{timeintervals} (A) illustrates this with the time interval $(t_0-w:t_0)$. Intuitively, as the utilization of the servers has a stochastic component, it is often the case that the servers will be accessed at different time intervals. Furthermore, different address locations of the DIMMs on the same server will be accessed at different time intervals. If certain cells are faulty, the sparsity in memory address accesses will lead to non-uniform temporal CEs distribution. By using fixed observation windows $(t_0-w:t_0)$, one may expect to miss details relevant for predicting memory failures. We found that there exist properties of failures that propagate through time through CEs that often surpass the observation window $(t_0-w:t_0)$ (e.g., the number of unique banks that generate CEs). Thereby, we consider structuring the CE reporting history into two disjoint time intervals, i.e., 1) $(-\inf:t_0-w)$ (Figure~\ref{timeintervals} (A) highlighted in grey) and 2) $(t_0-w:t_0)$ (the white rectangle). By proposing novel incrementally calculable features that among others characterize the difference between the disjoint intervals (e.g., the number of new banks generating correctable errors), we can preserve important information for upcoming failures. Motivated by our intuition, the focal point of this paper is to investigate the potential strengths of this view of the memory failure prediction problem. 

\section{Predicting Memory Failures with CEs}~\label{methodology}
In this section, we describe the procedure we implemented to learn a memory failure prediction model. It is a standard machine learning pipeline composed of data preprocessing, feature extraction and model learning. Once the model is learned it is used for online prediction. Figure~\ref{pipeline} (B) shows the pipeline. The input data comes in form of CE logs. They contain information for the DIMM name the CE appeared in, the timestamp, and the CE information (e.g., row, columns, banks, and similar). The log information proceeds towards the data preprocessing where CEs with incomplete log information are filtered. The filtered CEs proceed toward the feature extraction part. The latter implements a set of functions that characterize the degradation state of the DIMM. Once the features are extracted the CE represented with the feature vector is given as input for the machine learning model. In the following, we discuss each of the three parts in more detail.

\subsection{CEs Preprocessing and Feature Extraction}
The recorded CEs contain noise. For example, not all CE logs have information for the columns, rows or banks. Additionally, certain types of correctable errors, such as "uce.read" logs appear rarely. Therefore, the CEs preprocessing removes CEs of this kind (e.g., that are of type "uce.read", or have insufficient CE information). The filtered CEs proceeded toward the feature extraction part. 

The feature extraction part implements functions that characterize the degradation state of the DIMM. In total, we implemented 189 features calculable incrementally. To better describe them, we organize them in a taxonomy following DIMM properties and the failures are reflected. Figure~\ref{featuretaxonomy} gives the feature taxonomy. On the first level, there are five feature groups: \textit{general DIMM}, \textit{bank}, \textit{column}, \textit{row} and \textit{cell-level} features. The \textit{general DIMM} features characterize the overall CE statistics for the DIMM, irrespective of its components. This category includes features such as the \textit{frequency of errors}, or \textit{frequency} of the \textit{error types} similar as in related work~\cite{facebookmemoryfailureprediction}. We also proposed to use the relative change of the error frequency and error types between the two adjoint time intervals. The \textit{bank-level} features characterize the overall statistics of the CE frequency on a bank level, such as the number of banks that reported CEs, and the number of new banks that reported CEs between the observation window and the CEs before that. The focus of this category of feature is to describe the evolution of the failure on a bank level. 

The column, row and cell-level features characterized the DIMM banks on the memory array level. They are different in the way how they group the CEs alongside the memory array (e.g., by row, column or cell). For all the three groups there are three subgroups, i.e., \textit{neighbourhood}, \textit{repeating} and \textit{bank-row}(/column/cell) features. The \textit{repeating} group refers to the repetition of a CE from a single cell address or group of addresses sharing the same row/column. Intuitively, if one transistor is worn-off, any subsequent accesses to that memory will generate CEs. Therefore, repeating the same errors can indicate hard failures. The \textit{neighbourhood} features are similar to the repeating ones, however, they further exploit the concept of the DIMM \textit{bursting} mechanism. This concept enables reading out multiple adjacent memory addresses to speed up performance. Therefore, if an error occurs in the neighbourhood, it is likely that there are issues with the memory. The third group considers row/column/cell properties for the arrays irrespective of the bank and the chip. This group is closely related to the procedure of how the logical reading/writing of a whole word from multiple banks is conducted. 

\begin{table*}[!h]
\caption{Experimental Results}
\label{results}
\resizebox{\textwidth}{!}{
\begin{tabular}{|c|cc|ccc|ccc|ccc|}
\hline
\multirow{2}{*}{\begin{tabular}[c]{@{}c@{}}Feature\\ Calculation\\ Strategy\end{tabular}} & \multicolumn{2}{c|}{Method}                                                                                                          & \multicolumn{3}{c|}{RF}                                                                                                              & \multicolumn{3}{c|}{XGB}                                                                                                             & \multicolumn{3}{c|}{\begin{tabular}[c]{@{}c@{}}Operational\\ Practicies\end{tabular}}                                                                          \\ \cline{2-12} 
                                                                                                 & \multicolumn{1}{c|}{\begin{tabular}[c]{@{}c@{}}w\\ {[}h{]}\end{tabular}} & \begin{tabular}[c]{@{}c@{}}\# Normal\\ DIMMs\end{tabular} & \multicolumn{1}{c|}{Precision} & \multicolumn{1}{c|}{Recall}    & \begin{tabular}[c]{@{}c@{}}Error Rate\\ (Normal Test)\end{tabular} & \multicolumn{1}{c|}{Precision} & \multicolumn{1}{c|}{Recall}    & \begin{tabular}[c]{@{}c@{}}Error Rate\\ (Normal Test)\end{tabular} & \multicolumn{1}{c|}{Precision}              & \multicolumn{1}{c|}{Recall}                 & \begin{tabular}[c]{@{}c@{}}Error Rate\\ (Normal Test)\end{tabular} \\ \hline
\multirow{6}{*}{\begin{tabular}[c]{@{}c@{}}Overall \\ CE\\ Evolution\end{tabular}}               & \multicolumn{1}{c|}{\multirow{2}{*}{3h}}                                 & 2500                                                      & \multicolumn{1}{c|}{0.24±0.01} & \multicolumn{1}{c|}{0.58±0.01} & 0.27±0.01                                                          & \multicolumn{1}{c|}{0.58±0.01} & \multicolumn{1}{c|}{0.38±0.01} & 0.043±0.0                                                          & \multicolumn{1}{c|}{\multirow{2}{*}{0.279}} & \multicolumn{1}{c|}{\multirow{2}{*}{0.056}} & \multirow{2}{*}{0.027}                                             \\ \cline{3-9}
                                                                                                 & \multicolumn{1}{c|}{}                                                    & 5000                                                      & \multicolumn{1}{c|}{0.16±0.01} & \multicolumn{1}{c|}{0.49±0.01} & 0.20±0.01                                                          & \multicolumn{1}{c|}{0.48±0.01} & \multicolumn{1}{c|}{0.38±0.01} & 0.033±0.0                                                          & \multicolumn{1}{c|}{}                       & \multicolumn{1}{c|}{}                       &                                                                    \\ \cline{2-12} 
                                                                                                 & \multicolumn{1}{c|}{\multirow{2}{*}{168h}}                               & 2500                                                      & \multicolumn{1}{c|}{0.36±0.01} & \multicolumn{1}{c|}{0.55±0.01} & 0.15±0.01                                                          & \multicolumn{1}{c|}{0.60±0.01} & \multicolumn{1}{c|}{0.42±0.01} & 0.037±0.0                                                          & \multicolumn{1}{c|}{\multirow{2}{*}{0.109}} & \multicolumn{1}{c|}{\multirow{2}{*}{0.015}} & \multirow{2}{*}{0.006}                                             \\ \cline{3-9}
                                                                                                 & \multicolumn{1}{c|}{}                                                    & 5000                                                      & \multicolumn{1}{c|}{0.30±0.01} & \multicolumn{1}{c|}{0.46±0.01} & 0.09±0.01                                                          & \multicolumn{1}{c|}{0.47±0.01} & \multicolumn{1}{c|}{0.44±0.01} & 0.037±0.0                                                          & \multicolumn{1}{c|}{}                       & \multicolumn{1}{c|}{}                       &                                                                    \\ \cline{2-12} 
                                                                                                 & \multicolumn{1}{c|}{\multirow{2}{*}{336h}}                               & 2500                                                      & \multicolumn{1}{c|}{0.35±0.01} & \multicolumn{1}{c|}{0.54±0.01} & 0.14±0.01                                                          & \multicolumn{1}{c|}{0.61±0.01} & \multicolumn{1}{c|}{0.45±0.01} & 0.046±0.0                                                          & \multicolumn{1}{c|}{\multirow{2}{*}{0.31}}  & \multicolumn{1}{c|}{\multirow{2}{*}{0.06}}  & \multirow{2}{*}{0.011}                                             \\ \cline{3-9}
                                                                                                 & \multicolumn{1}{c|}{}                                                    & 5000                                                      & \multicolumn{1}{c|}{0.29±0.01} & \multicolumn{1}{c|}{0.45±0.01} & 0.07±0.01                                                          & \multicolumn{1}{c|}{0.48±0.01} & \multicolumn{1}{c|}{0.41±0.01} & 0.034±0.0                                                          & \multicolumn{1}{c|}{}                       & \multicolumn{1}{c|}{}                       &                                                                    \\ \hline
\multirow{6}{*}{\begin{tabular}[c]{@{}c@{}}Fixed\\ Window \\ Size\\ $[w]$\end{tabular}}              & \multicolumn{1}{c|}{\multirow{2}{*}{3h}}                                 & 2500                                                      & \multicolumn{1}{c|}{0.29±0.01} & \multicolumn{1}{c|}{0.48±0.01} & 0.17±0.01                                                          & \multicolumn{1}{c|}{0.37±0.01} & \multicolumn{1}{c|}{0.21±0.01} & 0.056±0.0                                                          & \multicolumn{1}{c|}{\multirow{2}{*}{0.23}}  & \multicolumn{1}{c|}{\multirow{2}{*}{0.023}} & \multirow{2}{*}{0.017}                                             \\ \cline{3-9}
                                                                                                 & \multicolumn{1}{c|}{}                                                    & 5000                                                      & \multicolumn{1}{c|}{0.20±0.01} & \multicolumn{1}{c|}{0.38±0.01} & 0.12±0.01                                                          & \multicolumn{1}{c|}{0.24±0.01} & \multicolumn{1}{c|}{0.23±0.01} & 0.055±0.0                                                          & \multicolumn{1}{c|}{}                       & \multicolumn{1}{c|}{}                       &                                                                    \\ \cline{2-12} 
                                                                                                 & \multicolumn{1}{c|}{\multirow{2}{*}{168h}}                               & 2500                                                      & \multicolumn{1}{c|}{0.31±0.01} & \multicolumn{1}{c|}{0.57±0.01} & 0.204±0.01                                                         & \multicolumn{1}{c|}{0.54±0.01} & \multicolumn{1}{c|}{0.38±0.01} & 0.05±0.0                                                           & \multicolumn{1}{c|}{\multirow{2}{*}{0.16}}  & \multicolumn{1}{c|}{\multirow{2}{*}{0.06}}  & \multirow{2}{*}{0.025}                                             \\ \cline{3-9}
                                                                                                 & \multicolumn{1}{c|}{}                                                    & 5000                                                      & \multicolumn{1}{c|}{0.24±0.01} & \multicolumn{1}{c|}{0.52±0.01} & 0.14±0.01                                                          & \multicolumn{1}{c|}{0.42±0.01} & \multicolumn{1}{c|}{0.43±0.01} & 0.047±0.0                                                          & \multicolumn{1}{c|}{}                       & \multicolumn{1}{c|}{}                       &                                                                    \\ \cline{2-12} 
                                                                                                 & \multicolumn{1}{c|}{\multirow{2}{*}{336h}}                               & 2500                                                      & \multicolumn{1}{c|}{0.31±0.01} & \multicolumn{1}{c|}{0.58±0.01} & 0.19±0.01                                                          & \multicolumn{1}{c|}{0.54±0.01} & \multicolumn{1}{c|}{0.35±0.01} & 0.051±0.0                                                          & \multicolumn{1}{c|}{\multirow{2}{*}{0.29}}  & \multicolumn{1}{c|}{\multirow{2}{*}{0.06}}  & \multirow{2}{*}{0.018}                                             \\ \cline{3-9}
                                                                                                 & \multicolumn{1}{c|}{}                                                    & 5000                                                      & \multicolumn{1}{c|}{0.22±0.02} & \multicolumn{1}{c|}{0.43±0.02} & 0.11±0.01                                                          & \multicolumn{1}{c|}{0.44±0.01} & \multicolumn{1}{c|}{0.37±0.01} & 0.036±0.0                                                          & \multicolumn{1}{c|}{}                       & \multicolumn{1}{c|}{}                       &                                                                    \\ \hline
\end{tabular}
}
\end{table*}

\subsection{Memory Failure Prediction}
As machine learning methods we consider two popular classification methods for tabular data, i.e., Random Forest~\cite{RF} and XGBoost~\cite{XGBoost}. These methods are often shown to achieve high performance on a plethora of tasks concerning tabular input representation~\cite{compariosn2017}. 

To learn the models, one requires labels for the individual CEs. As it is not known in advance the exact moment when the DIMM is in a degraded state, to label the data, we used a heuristic (similar as in related works~\cite{stefanovici2012cosmic}). We observed that the CEs for the failure DIMMs are usually generated in bursts that are time-delayed. In the cases when there are prior CE generated from a single DIMM, there is a large time between the adjacent bursts. To label the CEs representing the degraded DIMM state, we first find the largest existing gap between two adjacent CEs. This creates two sets of CE. We label all the CEs from the first set with 0. We consider the CEs from the second set to denote a degraded state and label them with 1. The correctable errors from new DIMMs in the current observation interval are given chronologically to the model. If the model predicts that at least one CE is degraded, the DIMM is predicted to fail. When learning the model, the CEs in the $m$ time units prior to the failure are removed. The final output of the system are the DIMM IDs, predicted to fail.

\section{Evaluation}~\label{evaluation}
\vspace{-6mm}
\subsection{Experimental Setup} 
To evaluate our methodology, we collected CE data with memory failures over six months (November-March 2020/21) from part of the server fleet of a large cloud provider. There were 12000 DIMMs with at least one CE. Around 3\% of the DIMMs failed. As the number of failed DIMMs composes 3\% of all the DIMMs, the problem of memory failure prediction indicates imbalanced classification. We adopted subsampling of the majority class (normal DIMMs), as a strategy to deal with this problem. To construct the training and test data, we split the normal DIMMs into two sets. We sampled 5000 normal DIMMs for cross-validation and used the remaining ones as a normal test set. We repeat the sampling five times and report the average results for the evaluation criteria. The validation normal DIMMs are paired with the failure DIMMs and are used to learn the model and access its performance. To access model performance we used 10 Fold CV. Precision and recall are common evaluation criteria for memory failure prediction~\cite{IBM2018}, and we adopt them. We set the value for $m$ to 3 hours, as it is sufficient time for server content migration.

For the window interval used to extract features, we used $w=\{3, 168, 332\}$ hours. The models were learned with the implementations of the sklearn-learn and xgboost Python libraries. Another baseline we consider is the thresholding of domain features (e.g., CE rate, or the number of uncorrectable errors) as frequent operational practices. Specifically, we experiment with all possible values for the CE rate and reported the best values for precision and recall.

\subsection{Results and Discussion}
Table~\ref{results} summarizes the results. When considering the overall CE evolution, one can observe that irrespective of the method or the experiment parameters, the results for both precision and recall are generally improved in comparison to those not considering it. For example, for a time window of 14 days and normal DIMMs 5000, for XGBoost the precision is improved for 9\% $(\frac{(0.48-0.44)}{0.44})$, while the recall for 11\% $(\frac{(0.41-0.37)}{0.37})$. Similar improvements of 32\% on precision and 5\% on recall are observed for RF for 14 days and normal DIMMs 5000. The average improvement in precision and recall for XGBoost is 33.6\% and 33\% accordingly, while for RF it is 8.1\% on precision and 5\% on recall. 

As seen by the results, including the long-term dependencies is particularly important for improving precision. This is likely related to the bursting nature of the CEs in anticipation of a failure. The long-term CE preservation enables the inclusion of non-related CEs from the normal DIMM behaviour that later failed. This effectively introduces information about the differences between the CEs of DIMMs that will fail in the future, guiding the model to better learn the normal DIMM states. This ultimately reduces the false positives and increases the precision. Another important point is that the model predictions outperform the manual operational practices. This is to the capabilities of the methods to combine complex information from multiple features. 

Finally, by close inspection of the features' importance, we noted that the most important features are the ones characterizing the transition between the two disjoint time intervals formed with the parameter $w$ (i.e., ($-\inf$, $t_0-w$) and ($t_0-w$, $t_0$)). This observation encourages further investigation into finding new features that better utilize the overall information. 

\section{Conclusion}~\label{conclusion}
In this paper, we revisited the problem of memory failure prediction with CEs. We found that considering the whole CEs-generation history is more indicative of failures, as compared to considering a fixed observation interval. We proposed a set of incrementally calculable features that preserve long-term CEs dependencies. In the evaluation of memory failures from the server fleet of a large cloud provider, we showed an average improvement of 21\% on precision and 19\% on recall, justifying the validity of our approach. This paper invites further research on how to better utilize the overall CEs data, to improve the performance on memory failure prediction. 
\bibliographystyle{IEEEtran}
\bibliography{IEEEabrv,referencesValid.bib}
\end{document}